\documentclass[twocolumn]{aa} 
\usepackage{graphicx}
\usepackage{txfonts}

\def\ergs{\mbox{ergs~s$^{-1}$}}
\def\sgra{\mbox{SGR~1806--20}}

\def\sgrc{\mbox{SGR~0526--66}}

\def\xte{XTE J1810$-$197}

\def\ee {1E\,1048.1$-$5937}
\def\kes {1E\,1841$-$045}

\def\ea {1E 2259$+$586}
\newcommand {\rc}{\rm}
\begin{document}
   \title{Discovery and Monitoring of the likely IR Counterpart of \sgra\
during the 2004 $\gamma$--ray burst-active state\thanks{The results reported in
this Letter are based on observations carried out at ESO, VLT, Chile
(programs 072.D--0297, 073.D--0381 and 274.D--5018)}}


   \author{GianLuca Israel\inst{1},  Stefano Covino\inst{2}, Roberto
Mignani\inst{3}, Luigi Stella\inst{1}, Gianni Marconi\inst{4}, Vincenzo
Testa\inst{1}, Sandro Mereghetti\inst{5},  Sergio Campana\inst{2}, Nanda
Rea\inst{1,6}, Diego G\"{o}tz\inst{5}, Rosalba Perna\inst{7} \and Gaspare Lo
Curto\inst{4}
          }

   \offprints{gianluca@mporzio.astro.it}

   \institute{INAF -- Osservatorio Astronomico di Roma, Via Frascati 
33,   I--00040 Monteporzio Catone (Roma),  Italy 
         \and
             INAF -- Osservatorio Astronomico di Brera, Via Bianchi 
46, I--23807  Merate (Lc), Italy
          \and 
	  	European Southern Observatory, Karl--Schwarzschildstr. 2,
D--85748 Garching, Germany
 	\and 
	European Southern Observatory, Casilla 19001, Santiago, Chile
	\and 
INAF -- Istituto di Astrofisica Spaziale e Fisica Cosmica ''G.Occhialini'',   
Via Bassini 15, I--20133 Milano, Italy	
	\and
SRON -- National Institute for Space Research, Sorbonnelaan 2,
3584 CA, Utrecht, The Netherlands
          \and
Department of Astrophysical and Planetary Sciences and JILA, University
of Colorado, 440 UCB, Boulder, CO, 80309, USA	     }

   \date{Received 6 April 2005 / Accepted 3 June 2005}

\abstract{The sky region including the Chandra position of \sgra\ was monitored
in the IR band during 2004,  following its increased high energy bursting
activity. Observations were performed using NAOS-CONICA, the adaptive optics IR
camera mounted on Yepun VLT,  which provided images of unprecedented quality
(FWHM better than 0\farcs1). After the 2004 December 27th giant flare, the
source position has been nailed by VLA observations of its radio counterpart,
reducing the positional uncertainty to 0\farcs04.   Using IR data from our
monitoring campaign, we discovered the {\rc likely} IR counterpart
to \sgra\ based on positional coincidence with the Chandra and VLA uncertainty
regions and flux variability of a factor of about 2 correlated with that at
higher energies. We compare our findings with other isolated neutron star
classes thought to be related, at some level, with SGRs.
   }
   \titlerunning{Discovery and monitoring of the likely IR counterpart to \sgra}
   \authorrunning{Israel et al.}
   \maketitle
%

\section{Introduction}
Soft Gamma--ray Repeaters (SGRs) were discovered in the seventies through the
detection of short ($<$1s), recurrent, and  intense bursts of high energy
emission peaked in the soft $\gamma$ rays. 
Only four  confirmed SGRs are known, three in the Galaxy and one in the Large
Magellanic Cloud (for a review see e.g. Woods \& Thompson 2004).   The detection
of a $\sim 8$~s periodicity in the decaying tail of a very intense
($\sim$10$^{44}$\,ergs) and long (several minutes)  event, known as
giant flare,  from \sgrc\ on  1979 March 5th (Mazets et al. 1979) suggested the
association of SGRs  with  neutron stars.
A small sample of peculiar X--ray pulsars, namely the Anomalous X--ray Pulsars
(AXPs) has been proposed to be closely related to SGRs based on 
similar properties, namely their period P (in the 5--12\,s range),  their period
derivative \.P (10$^{-10}$--10$^{-13}$ s\,s$^{-1}$ range), and  X--ray bursts
(Kouveliotou et al. 1998;  Kaspi et al. 2003; Gavriil et al. 2002). 

Both SGRs and AXPs have been proposed to be powered by the decay of  strong
magnetic fields that characterise these neutron stars (B $\sim$
10$^{14}$--10$^{15}$ G; Duncan \& Thompson 1992; Thompson \& Duncan 1995). The
``magnetar'' model is founded on two observational facts: firstly, the
rotational energy loss inferred from the SGR and AXP spin--down is insufficient
to power their persistent X--ray luminosity of $\sim$10$^{34}$--10$^{36}\ergs$;
secondly, there is no evidence for a companion stars which could provide the
mass to power the X--ray emission through accretion.  
   \begin{table}
      \caption[]{Journal of the VLT NACO IR 2004 observations for the field of
\sgra. 
}
         \label{LogObs}
         $$
	 \begin{array}{clccl}
            \hline
            \noalign{\smallskip}
            2004~ Start~ UT   & Filter & Exposure & FHWM & {\rc Ks ~{\rc mag}}\\
            ({\rm MM~DD~/~HH:MM} ) & & ({\rm s})  &  (\arcsec) & {\rc for
~object ~A}
\\
	    \noalign{\smallskip}
            \hline
            \noalign{\smallskip}
03~ 10 ~/~09:00 & Ks& 1200 &0.14 	& {\rc contaminated} \\
03~ 17 ~/~08:14 & J& 1980 &0.21	&	{\rm -} \\
03~ 17 ~/~08:53 & Ks& 1800 &0.10	&	{\rc 20.01\pm0.14} \\
03~ 17 ~/~09:33& H& 1800 &0.14	&	{\rm -} \\
03~ 18 ~/~07:35 & Ks& 3600 &0.09	&	{\rc 20.09\pm0.15} \\
06~ 13 ~/~03:38 & Ks& 3600 &0.11	&	{\rc 20.26\pm0.26}\\
06~ 19 ~/~04:51 & Ks& 6120 &0.11	&	{\rc 19.94\pm0.13}\\
08~ 10 ~/~02:50 & Ks& 2640 &0.12	& 	{\rc contaminated}	\\
08~ 11 ~/~03:04 & Ks& 5160 &0.10	&	{\rc 19.48\pm0.12}\\
09~ 07 ~/~23:56 &Ks& 5240 &0.09	&	{\rc 19.70\pm0.08}\\
10~ 05 ~/~23:45 & Ks& 5160 &0.11	&	{\rc 19.32\pm0.16}\\
            \hline
         \end{array}
     $$ 
   \end{table}

Bursting activity from \sgra\ resumed at the end of 2003 displaying an increase
in both the $\gamma$--ray burst rate and the hard X--ray persistent emission
(Mereghetti et al. 2005a) throughout 2004, and culminating with the giant flare
of  27th December 2004 (Borkowski et al. 2004), during which
$\sim$10$^{47}$\,ergs were released ({\rc for a distance of about 10kpc; Cameron
et al. 2005; McClure-Griffiths  \& Gaensler 2005}). Few days after this event,
\sgra\ was observed and detected in the radio  band for the first time,
providing very   accurate positions (VLA; Cameron et al. 2005; Gaensler et al.
2005). 

In this work we report on the results of an extended Target of Opportunity (ToO)
observational campaign on \sgra\ carried out during 2004 with  the ESO VLT. 
In particular, we report on the {\rc likely} discovery of the IR counterpart to
\sgra\ based on positional coincidence with the radio and Chandra positions and
flux variability.  (Preliminary results were reported in Israel et al. 2004,
2005a, 2005b, before and independently from Kosugi et al. 2005.) 
We briefly compare the IR emission properties of \sgra\ with those of related
objects.


\section{NAOS-CONICA observations at VLT}
   \begin{figure*}
   \includegraphics[width=9.2cm]{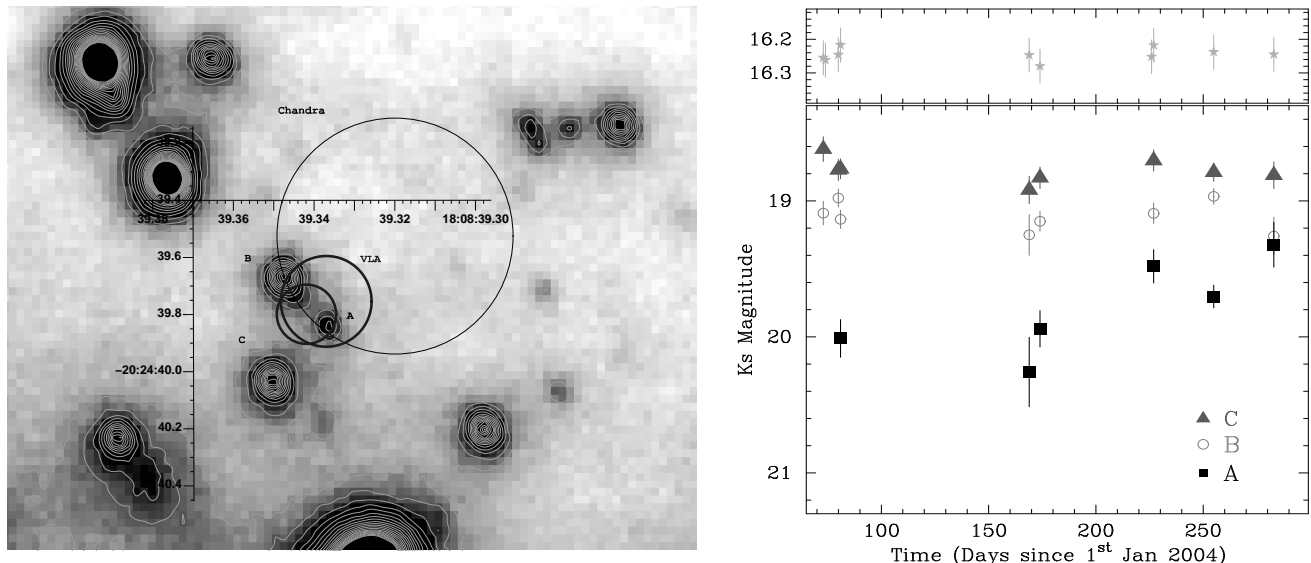}
   \caption{
IR NACO Ks band image (left panel; 18th June 2004 dataset) close--up of the
$1\farcs5\times 1\farcs5$  portion of the sky around the {\rc 1$\sigma$} Chandra
and VLA uncertainty circles (radius of 0\farcs3, 0\farcs04 and 0\farcs1,
respectively) with the proposed counterpart marked with A.  Coordinates are RA
(h\,m\,s) and Dec. ($^{\rm o}$ \arcmin\ \arcsec; equinox J2000). Isophotal
contour lines are also drawn for clarity. Ks light curves (right panel) for a
number of selected sources:  the {\rc proposed} counterpart to \sgra\ ($A$) with
two nearby objects (lower panel), and the closest reference (isolated and
bright) star (upper panel) we used for relative photometry among Ks frames. Flux
variability is clearly detected only for object $A$.}
              \label{Figcircles}%
    \end{figure*}

The observations presented here were performed as part of an ESO Target of
Opportunity program extending over October 2003 -- September 2004, and a
Director Discretionary Time observation on October 2004. Observations were
triggered following the  detection of intense $\gamma$--ray bursts or during
epochs of increased burst rate (G{\"o}tz et al. 2004; Golenetskii et al.
2004a,2004b). The data were acquired at Yepun VLT with the Nasmyth Adaptive
Optics System and the High Resolution Near IR Camera providing a pixel size of
0\farcs027 (NAOS-CONICA; see Table\,\ref{LogObs} for details).  
For all observations we used an exposure time of 40\,s and a number of frames
per image of 3 with a random offset of 7\arcsec\ among images in order to
perform background subtraction of the variable IR sky. VLT NACO science images
were reduced based on the standard tools provided by the ESO - Eclipse package
(Devillard 1997).  As a result of the presence of the Ks=8.9 magnitude object
LBV 1806--20 (see Eikenberry et al. 2004) close to the edge of the NACO field of
view (FOV),  artificial ring--like ghost structures were clearly detected in the
image at coordinates R.A.=18$^{\rm h}$ 08$^{\rm m}$ 40\fs31;
Dec.=$-$20$^{\circ}$ 24\arcmin\ 41\farcs21 (equinox 2000),  13\arcsec\  away
from the LBV 1806--20 position.  In order to reduce as much as possible the
effects of contamination due to nearby objects, relative aperture and Point
Spread Function (PSF) photometry was obtained within narrow annuli (about 1--1.5
FWHM depending on the seeing conditions), while the background was evaluated
close to the object under analysis. Absolute photometry was derived by analysis
of the best seeing frames.
Finally, we cross--checked  our absolute magnitudes by means of archival ISAAC 
data of the same region and about $100$ isolated stars taken from the 2MASS
catalog and within the instrument FOV: the results were in agreement to within
0.05 Ks magnitudes. 
 
In order to register the Chandra and VLA  coordinates of \sgra\ on our IR
images, we obtained the image astrometry by using the positions of
about 10 stars selected from the 2MASS catalogs and within the 
$\sim$30$\arcsec  ~\times$ 30$\arcsec$ NACO FOV of final images.  The residual
in the fit was of 0\farcs06 in each coordinate, converting to $\sim 0\farcs1$
once the  2MASS absolute accuracy was
included\footnote{http://www.ipac.caltech.edu/2mass/releases/allsky/doc}.
\rc Fig.\,\ref{Figcircles} shows the $\sim$\,1\farcs5$\times$1\farcs5  Ks band
region around the Chandra and VLA positions (1$\sigma$ confidence level radius
of 0\farcs3, 0\farcs04, respectively; Kaplan et al. 2002; Gaensler et al.
2005). However, given that the Gaensler et al. (2005) radio position refers to
about 20 days after the giant flare of \sgra, and that the  source from which 
is originating the radio emission is moving at about 4\,mas/day (Taylor et al.
2005), we  also plot the VLA position obtained after 7 days by Cameron et al.
(2005; 1$\sigma$ radius if 0\farcs1), corrected for about 30\,mas in right
ascension (following Taylor et al. 2005);
this corresponds to a final 1$\sigma$ confidence level radius of 0\farcs14. 
Source $A$, a relatively faint (Ks$\sim$20) object, at the sky position R.A.=
18$^{\rm h}$ 08$^{\rm m}$ 39\fs337, Dec.= $-$20$^{\circ}$ 24\arcmin\ 39\farcs85
(equinox 2000, 90\% uncertainty of 0\farcs06 ), is found to be consistent with
the Chandra and VLA positional uncertainty circles superimposed on our IR
astrometry--corrected frame. Objects $B$ and $C$ ($\sim$0\farcs23 and 0\farcs27
away from $A$, respectively; by looking at the contour lines, we note
that  object $B$ might be the blend of two unresolved objects) are only
marginally consistent with the X--ray and radio positions, even though
statistically plausible. Object $A$ was not detected in the J and H images;
3$\sigma$ upper limits of magnitude 21.2 and 19.5 were derived, respectively. 
We note that the \sgra\ IR counterpart $A$ {\rc (Ks magnitudes are listed in
Table\,\ref{LogObs})} plus objects $B$ and $C$ (Ks=19.07$\pm$0.04 and
18.77$\pm$0.04, respectively) are all within a radius of $\sim$0\farcs25 from
the VLA positions, and consistent with being unresolved components of candidate
$B$ in the IR images of Eikenberry et al. (2001; K=18.6$\pm$1.0).  

Light curves of the $A$, $B$ and $C$ objects marked in Fig.\,\ref{Figcircles}
are shown in Fig.\,\ref{Figcircles} (right plot). Candidate $A$ is the only one
showing a clear brightening (a factor of $\sim$2) in the IR flux between June
and October 2004. Objects  $B$ and $C$ show a fairly constant flux\footnote{\rc
For the faintest component of blended object B we can reasonably exclude any IR
variability, similar to that shown by $A$, for any Ks magnitude
brighter than about 22.5 .}. The upper panel of Fig.\,\ref{Figcircles} (right
plot) shows the closest reference star (1\farcs6 away form the target) used for
relative photometry across Ks images: the object is constant to within the
photometric uncertainties. We thus conclude that object $A$  is variable.
We checked for a similar variability also in  the X--ray flux of \sgra. Both
the XMM--Newton (Mereghetti et al. 2005b) and INTEGRAL (Mereghetti et al. 2005a)
persistent fluxes of \sgra\ showed an increase across the two semesters of 2004
by  a factor of 1.94$_{-0.02}^{+0.01}$ and 1.7$_{-0.3}^{+0.4}$ in the
2--10\,keV and 20--100\,keV bands, respectively\footnote{for the 2--10\,keV 2004
first semester flux  we assumed that of October 2003, based on the unvaried
INTEGRAL flux  between October 2003 and February--April 2004.}. During the same
time interval the NACO Ks flux increased by a factor of 2.4$_{-0.5}^{+0.9}$,
consistent with high energy flux variations. This further supports the
identification of object A as the correct IR counterpart of \sgra. 

Recently, independent from our work, the object $A$ has
been proposed as the  IR counterpart to \sgra\ (Kosugi et al. 2005; their object
$B3$). A comparison of their photometry with our we shows that nearly all the Ks
magnitudes have an  offset of about 0.2, with the important exception of objects
$A$ and $C$ which are 1.6 and 0.6 Ks magnitudes brighter than the corresponding
objects $B3$ and $B1$ in Kosugi et al. (2005), respectively.  Even
though we do not have a clear explanation for the observed differences, we note
that in our images we did not see any evidence for (i) a brightening of objects
$B$ and $C$, and (ii) an increase of the local background around object $A$
(based on our best datasets with FWHM $\leq$0\farcs1), in contrast to Kosugi et
al. (2005).  An unusually high background level (regardless of its
origin) may of course result in a flux underestimation of a source that lies in
the same area.  

\section{Discussion}
The deep and high spatial resolution NACO images allowed us to 
identify the likely IR counterpart of \sgra, and monitor its IR flux
for seven months in 2004, during which an increase by a factor of $\sim$2
was detected, correlated with the flux in the high energy bands.
In fact, the IR flux of \sgra\  was fairly constant until mid-June 2004,  
while it grew rapidly between June (Ks=20.01$\pm$0.14) and October 2004
(Ks=19.32$\pm$0.16; 1$\sigma$ uncertainties; these values override the
preliminary ones reported in Israel et al. 2005b). 

IR variability has been 
detected in nearly all AXPs with known IR counterpart. In particular, for
\ee, \xte\ and \ea, IR variability has been found, or suspected, to be
correlated with the persistent X--ray emission (Israel et al. 2002; Rea et al.
2004; Tam et al. 2004).  Based on the NACO results we can conclude that the
IR/X--ray correlation observed in AXPs also holds for \sgra. 
The total fluence of the IR
enhancement between June and October 2004 is  about $10^{41}$\,ergs (we
assumed $A_V$=29$\pm$2; see Eikenberry et al. 2004), a
factor of about 100 smaller than that in the 2--10\,keV band. 
   \begin{figure}
   \centering
   \includegraphics[angle=-90,width=\columnwidth,clip]{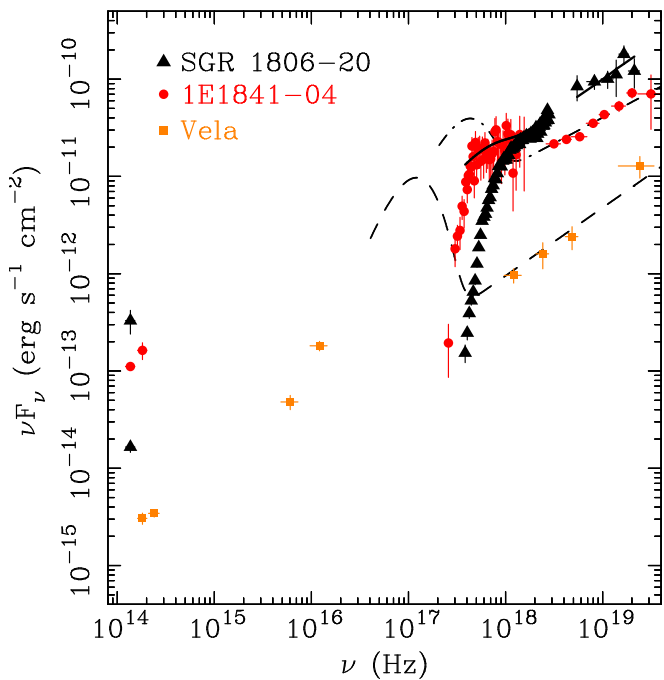}
      \caption{Broad band energy spectrum of \sgra\ (triangles) and, as a
comparison,  the AXP \kes\ (circles) and the radio pulsar Vela (squares).  
In the case of \sgra\  and \kes\ high energy data are taken from 
XMM--Newton (Mereghetti et al. 2005b; referring to 6th October 2004 for \sgra),
Chandra (for \kes) and INTEGRAL (Mereghetti et al. 2005a; 21st September -- 14th
October 2004 for \sgra). Absorbed and unabsorbed IR fluxes ($A_{\rm
V}$=29$\pm2$, 5th October 2004 NACO observation) are shown in the case
of \sgra, unabsorbed ($A_{\rm V}$=13$\pm1$)  IR fluxes are instead
reported for the likely candidate of \kes\ (circles; Israel
et al. 2005, in preparation). All the data for Vela are taken form literature
(Kaspi, Roberts \& Hardings 2004 and references therein). Solid curves
(continuous, stepped and dot--stepped) are the unabsorbed fluxes for the black
body plus power law model used to fit the high energy part of spectra.}
         \label{v_vfv}
   \end{figure}
%

Based on the above reported findings we note that the \sgra\ emission
varies in a similar fashion (in terms of timescale and amplitude of variation)
over more than five orders of magnitude in photon energy. The similar flux
variation in the IR and X--ray bands suggests that the emission in the two bands
has a similar, if not the same, origin. Moreover, it has become evident that
{\rc X--ray flux enhancement of the persistent emission of
SGRs  is correlated with their burst rate, making it difficult to compare the 
fluxes among different SGRs without knowing their burst history}  (see Woods \&
Thompson 2004). Tam et al. (2004) argued that IR thermal surface emission
(within the magnetar model) is ruled out during the correlated X--ray/IR flux
decay phases of \ea\ (implausibly high implied brightness temperature),
suggesting the magnetospheric origin for the IR enhancement.  {\rc
Alternatively, the IR flux can be due to reradiation by material in the vicinity
of the the pulsar. This model naturally predicts a correlation between the the
IR and the X--ray flux (Perna, Hernquist \& Narayan 2000; Rea et al. 2004).}

{\rc This is the first time that the broad band energy properties of an SGR can
be compared, over a similar energy band, with those of other classes of isolated
neutron stars, such as AXPs and radio pulsars.}
In Fig.\,\ref{v_vfv} we show the ``nearly simultaneous'' broad band energy
spectrum of \sgra\ from the IR to $\gamma$ rays (high energy data are taken form
Mereghetti et al. 2005a; see caption for details). The high energy part of the
spectrum is clearly consistent with being non--thermal emission (a power--law
model is generally used)  from the source. We also plot the
spectrum from the AXP \kes, for which 20--200\,keV band data are available
(Kuiper et al. 2004); a similar non--thermal component is displayed by the
source.  Non--thermal components are also seen in radio pulsars and modelled 
with power--law components (see Kaspi, Roberts \& Harding 2004 for a recent
review). In some cases there is a smooth connection between optical, X--rays and
$\gamma$--ray emission (Crab), while in other cases the extrapolation is
plausible (Vela; see Fig.\,\ref{v_vfv}).  {It is worth noting the similar flux
ratios in the IR and hard X--ray bands for the three objects}, and the
significant difference of the characteristic temperature of thermal soft X--ray
components between radio pulsars ($\leq$0.1\,keV) and SGRs/AXPs
(0.4--0.8 keV for a BB fit and 0.2--0.5 keV for a magnetic atmosphere fit,
Perna et al. 2001), suggesting a significantly larger energy injection on the
neutron star surface in ``magnetar'' candidates than in radio pulsars.

Future detailed multi--wavelength observations campaigns of AXPs and SGRs will
likely help  clarifying the link between IR and high energy bands.  Furthermore,
the detection of the quiescent IR flux level of \sgra\ will allow to compare
the net energy released by the source in the IR and X--ray/$\gamma$--ray bands
during its bursting active phase.

\begin{acknowledgements}
We thank the ESO Director's Discretionary Time Committee for accepting 
the observation  of \sgra\ few hours after the 5th October 2004 intense X--ray
burst. We are also indebted with  VLT personnel for their continuous help in
optimising and  performing the NACO observations.
We thanks D. Dobrzycka and W. Hummel for their help in clarifying the artificial
nature of the rings in the NACO images. This work was partially supported
through Agenzia Spaziale Italiana (ASI), Ministero  dell'Istruzione,
Universit\`a e Ricerca Scientifica e Tecnologica (MIUR -- COFIN), and Istituto
Nazionale di Astrofisica (INAF) grants. N.R. is supported by a Marie Curie
Trainig Grant  (HPMT-CT-2001-00245).
\end{acknowledgements}

\end{document}